\documentclass[aps,prl,twocolumn]{revtex4}
\usepackage{graphicx}
\usepackage{psfrag}
\begin{document}
\newcommand{\p}{{\partial}}
\newcommand{\pr}{{\prime}}
\newcommand{\beq}{\begin{equation}}
\newcommand{\eeq}{\end{equation}}
\newcommand{\bqa}{\begin{eqnarray}}
\newcommand{\eqa}{\end{eqnarray}}
\newcommand{\g}{{\nabla}}
\newcommand{\bn}{{\mathbf n}}
\newcommand{\bu}{{\mathbf u}}
\newcommand{\rd}{{\rm d}}

\title{Viscous Withdrawal of Miscible Liquid Layers}
\author{Laura E. Schmidt and Wendy W. Zhang}
\affiliation{The Department of Physics \& The James Franck Institute, 
University of Chicago, Chicago, Illinois 60637, USA}
\date{\today}

\begin{abstract}
In viscous withdrawal, a converging flow imposed in an upper layer 
of viscous liquid entrains liquid from a lower, stably stratified 
layer.  Using the idea that a thin tendril is entrained by a local 
straining flow, we propose a scaling law for the volume flux of 
liquid entrained from miscible liquid layers.
%Using the idea that the liquid is entrained in a thin tendril
%by a local straining flow, we propose a scaling law for the volume flux 
%entrained from miscible liquid layers.  
A long-wavelength model 
including only local information about the withdrawal flow is 
degenerate, with multiple tendril solutions for one withdrawal 
condition.  Including information about the global geometry of the 
withdrawal flow removes the degeneracy while introducing only a 
logarithmic dependence on the global flow parameters into the 
scaling law.  
\end{abstract}
\maketitle
\begin{figure}
%\vspace{-10pt}
\includegraphics[width=0.3\textwidth, angle=0]{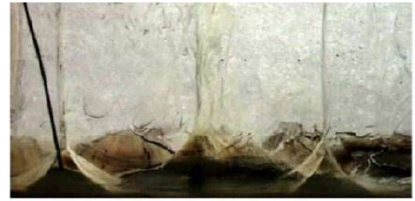}
\vspace{-5pt}
\caption{\label{exp}Axisymmetric tendrils entrained by a warmer, 
upwards flow in a 2-layer thermal convection experiment with 
miscible oils. Convection is driven by heating the bottom of 
the tank and cooling the top. The lower fluid is less viscous 
than the upper and the spacing between tendrils is $O$($10$ cm).  
The dark vertical line on the left is a thermocouple~\cite{jellinek04}.}
\vspace{-15pt}
\end{figure}

Recent experiments on thermal convection with two layers of 
miscible liquids reveal several distinct regimes~-- an overturn 
regime with violent mixing of the layers, a doming regime where 
the interface undulates, and a stratified regime where the 
convection is largely stable with thin tendrils or sheets of one 
liquid entrained within the 
other~\cite{davaille99b}.  
%other~\cite{davaille99b, davaille04}.  
Analogous steady-state entrained structures arise in drainage 
flows~\cite{jeong92,courrech06}, 
oil extraction~\cite{blake86}, 
%oil extraction~\cite{blake86,lister89}, 
as well as viscous withdrawal of immiscible liquid layers, which occur in 
microfluidics~\cite{ganan-calvo04}, 
%microfluidics~\cite{suryo06,anna03,ganan-calvo04}, 
fiber coating~\cite{simpkins00} 
%fiber coating~\cite{simpkins00,egg01,lorenceau03,lorenceau04} 
and encapsulation of biological 
cells~\cite{cohen01}.  
%cells~\cite{cohen01,wyman07}.  
Recent works exploring the connections between thermodynamic 
phase transitions and the topology transition that takes place 
at the onset of entrainment have noted that, in order for the 
entrained structure to be completely isolated from the 
large-scale flow dynamics, the shape of its base must be a 
power-law cusp~\cite{case07, wz04, courrech06, cohen02}.  
Intriguingly, experiments~\cite{davaille02, manga02} on miscible 
entrainment also seem to show a robust cusp-like shape at the 
base of long-lived tendrils (see Fig.~\ref{exp}).  This 
suggests the entrained tendrils are isolated from the 
fluctuating, large-scale convection by the cusp-shaped base and 
are therefore able to remain stable over many convection cycles.  
If true, this may even explain why hot-spots can persist over 
many convection cycles in the Earth's mantle~\cite{davaille02,jellinek04}.  
Motivated by these observations, we focus on the stratified 
regime in thermal convection of miscible layers and present a 
model that tests how the large-scale flow and topography anchor 
a thin cylindrical tendril.

Because the large-scale flow is stabilized in the stratified 
regime, mixing between the layers is controlled by the volume 
flux of liquid entrained through the tendrils, $Q_0$. Existing 
estimates of $Q_0$ assume that the velocity field inside the 
tendril is uniformly upwards, flowing at a characteristic convection
speed~\cite{sleep88,davaille02, manga02}.  
%with the same speed as 
%the overall convection~\cite{sleep88,davaille02, manga02}.  
However, recent particle-image-velocimetry (PIV) measurements 
within the base of an anchored tendril reveal a stagnation-point 
velocity field, one more appropriately described by a 
characteristic strain rate $E$ (s$^{-1}$) instead of a 
characteristic velocity scale~\cite{davaille_focusing}.  
Viscous withdrawal experiments on immiscible layers suggest how 
an interior stagnation point can arise~\cite{case07}.  When the 
effect of the entrainment penetrates deeply into the lower layer, 
a broad tendril forms with the interior moving uniformly upwards. 
This is the situation addressed by  existing estimates. When the 
effect of the entrainment penetrates weakly, a narrow tendril 
forms by drawing liquid inward within a thin layer below the 
interface, thus creating an interior stagnation point.  Here we 
derive a new scaling law for the volume flux entrained, one which 
takes into account the interior stagnation point. 

\begin{figure}
\includegraphics[width=0.4\textwidth, angle=0]{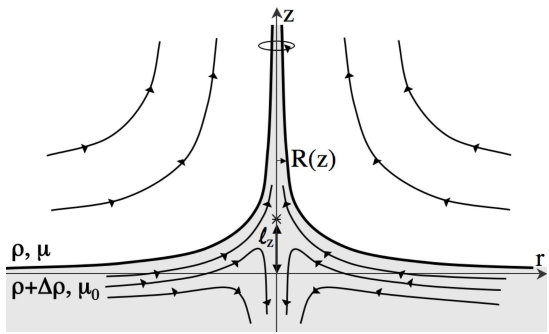}
\vspace{-8pt}
\caption{\label{setup} Sketch of surface $R(z)$ and flow streamlines for an 
  axisymmetric withdrawal flow in the upper liquid layer entraining 
  a thin, cylindrical tendril from a deep lower layer.}  
\vspace{-12pt}
\end{figure}

%straining flow vs. uniform flow
The steady-state flow fields associated with non-linear interface 
deformations are difficult to obtain analytically. We therefore 
analyze the simplest case: the  steady-state entrainment of a 
tendril from a deep lower layer by an axisymmetric viscous 
withdrawal flow imposed in an upper layer (Fig.~\ref{setup}). 
The entrained liquid is taken to be much less viscous than the 
exterior, so that there is only a weak feedback from entrained 
flow to the withdrawal flow.  To ensure that the effect of 
entrainment penetrates only weakly into the lower layer, we 
require that the layers are strongly stratified. With these 
simplifications, the steady-state interface divides into three 
geometrically distinct regions: a long and slender entrained 
tendril where interior flow effects are dominant, a far-field 
interface where the hydrostatic pressure is dominant, and a 
transition region where the force balance changes smoothly from 
one form to another. 

First we will give a scaling argument that predicts the volume 
flux $Q_0$. The system is best described by a cylindrical coordinate 
system where $z=0$ corresponds to the height of the undisturbed 
interface and $r=0$ corresponds to the centerline of the tendril. 
The steady-state tendril radius is defined as $R(z)$. In the 
strongly stratified regime considered here, the stagnation point inside the 
tendril lies at a height $\ell_z$.  The velocity field near the 
stagnation point is an axisymmetric straining flow,
$(-Er/2)\, {\bf e}_r + E z\, {\bf e}_z$. To estimate $Q_0$ we note 
that, above $\ell_z$, the liquid from the lower layer is moving 
entirely upwards so that $Q_0 \approx (Ez)\pi R^2(z)$.  
Since it is possible to relate $\ell_R$, the tendril radius
near the stagnation point, to $\ell_z$ in a simple way described below,
the volume flux is $Q_0 \approx (E \ell_z) \pi \ell^2_R$.
%We also note that it is possible to estimate $\ell_R$, the tendril 
%radius near the stagnation point, in terms of $\ell_z$ in a simple 
%way described below.   Given $\ell_R$ and $\ell_z$, the volume 
%flux is $Q_0 \approx (E \ell_z) \pi \ell^2_R$.  

Assuming the tendril is long and slender in the neighborhood of 
the stagnation point, the interior velocity field can be approximated 
%we can approximate the interior velocity field
via standard slender body type arguments~\cite{acrivos78}. 
The upwards contribution to the interior flow due to entrainment 
is then a uniform plug flow, $Ez$.  The downward contribution is 
a pipe flow driven by hydrostatic pressure $\Delta \rho g 
(R^2(z)-r^2)/(4 \mu_0)$, where $\Delta \rho$ is the density 
difference between the two layers, $g$ the gravitational 
acceleration,  and $\mu_0$ is the viscosity of the entrained 
liquid.  By definition, these two flows must cancel at the 
interior stagnation point ($r=0, z=\ell_z$) so that
\beq
E \ell_z \sim \frac{\Delta \rho g}{4 \mu_0} \ell^2_R .
\label{ratio}
\eeq
It remains to determine $\ell_z$.  We identify $\ell_z$ 
as the amplitude of the large-scale upwards deflection produced 
on the interface by the withdrawal flow. It is therefore set by 
a balance of $\mu E$, the upwards pull exerted by viscous 
stresses in the upper-layer withdrawal flow, where $\mu$ is 
the viscosity of the liquid in the upper layer, and $\Delta \rho g$, 
the downwards pull of hydrostatic pressure.  To simplify the 
notation for later analysis which uses $\ell_z$ as a 
characteristic lengthscale, we define
\beq
\ell_z \equiv {2 \mu E}/{\Delta \rho g}.
\label{lz}
\eeq
Relations (\ref{ratio}) and (\ref{lz}) together yield
\beq
\ell_R = \sqrt{ 2 \left( \frac{2 \mu_0 E}{\Delta \rho g} \right) \left( \frac{2 \mu E}{\Delta \rho g} \right) }, 
\label{lr}
\eeq
corresponding to the geometric mean of the viscous lengthscale 
associated with the upper layer flow and the, much smaller, 
viscous lengthscale associated with the less viscous, lower 
layer flow.  The tendril ``slenderness ratio'', $\ell_R/\ell_z$, 
assumed to be small at the beginning of our analysis, is 
$\sqrt{2\mu_0/\mu}$, which is indeed small for $\mu_0 \ll \mu$. 

Finally, using (\ref{ratio}) and (\ref{lz}), we can rewrite $Q_0$ in 
terms of the local strain rate $E$ and material parameters
\beq
Q_0 = 16 \pi c_0 {\mu^2 \mu_0 E^4}/{(\Delta \rho g)^3},
\label{fluxlaw}
\eeq
where $c_0$ is an $O(1)$ dimensionless entrainment coefficient
which we determine later in the full analysis.  This scaling 
law differs from existing estimates based on a characteristic 
convection velocity $U$~\cite{sleep88,davaille02, manga02}. 
If we take $S$ as the size of the convection cell, then 
(\ref{fluxlaw}) says $Q_0 \propto (U/S)^4$ while existing 
estimates say $Q_0 \propto U^3$.  We also emphasize that 
nothing in the above scaling argument depends on the withdrawal being 
axisymmetric. An analogous argument for entrained 2-d sheets 
yields the same relations for $\ell_z$ (\ref{lz}) and the 
sheet thickness $\ell_R$ (\ref{lr}), and the scaling law 
$Q_0/L\approx E \ell_z \ell_R$ for the volume flux per unit length.

In deriving the scaling law for $Q_0$ (\ref{fluxlaw}) we 
essentially assumed the entrainment dynamics is controlled 
by a local straining flow.  Since a straining flow has no 
inherent lengthscale, the dynamics is completely decoupled 
from the geometry of the large-scale withdrawal flow.  In 
thermal convection, this feature would imply that the volume 
flux $Q_0$ has no dependence on the size of the large-scale 
convection cell. The last implication is counter-intuitive.  
After all, the tendril size at its base, where it joins on 
the interface, is controlled by how the interface levels out 
and becomes flat on the large lengthscales. This large-scale 
topography in turn depends on the specific geometry of the 
withdrawal flow, so physical intuition says that the 
entrainment dynamics, particularly $Q_0$, should also depend 
on the global geometry. 

We next address this question by developing a long-wavelength 
model of the entrainment dynamics and find that a model which 
includes only the local straining flow is degenerate. For each 
value of $Q_0$, there exists a continuous family of tendril 
solutions, each with a different power-law shape at the base.  
The existence of a steady-state tendril thus does not require a 
specific power-law base shape. Including the large-scale 
withdrawal flow geometry removes this degeneracy. It also 
introduces a logarithmic dependence on the global dynamics 
into the scaling law for $Q_0$ (\ref{fluxlaw}).  

Starting with the Navier-Stokes equations, and using standard 
slender-body approximations where effects proportional to 
$\ell_R/\ell_z$ are discarded at leading-order, we derive the 
following equation for the steady-state tendril shape
\beq
Q_0 = (Ez) \pi R^{2}(z) - \frac{\pi R^{4}(z)}{8 \mu_0}\frac{\rd P_0}{\rd z}
\label{goveqn}
\eeq
and interior pressure
\beq
P_0 = 2 \mu E \left( 1 + \frac{z}{R(z)} \frac{\rd R}{\rd z} \right) +\Delta\rho g z. 
\label{pressure}
\eeq
In essence, equation (\ref{goveqn}) says that the {\it unknown} 
volume flux of liquid entrained into the tendril has two 
contributions: a plug flow $Ez$ induced by the withdrawal, and 
a pipe flow associated with non-uniform interior pressure. 
The expression for $P_0$ (\ref{pressure}) derives from the 
normal stress balance across the surface of the tendril.  The 
second term is the familiar hydrostatic pressure difference. 
The first term is the large pressure necessary to keep a 
tendril from collapsing under the inward squeeze exerted by 
the exterior straining flow.  Equations (\ref{goveqn}) and 
(\ref{pressure}) together yield a second-order nonlinear 
ordinary differential equation for the steady-state tendril 
shape $R(z)$ and $Q_0$.  The derivation is analogous 
%to that employed in drop 
%emulsification studies~\cite{acrivos78, eggleton01}.
%emulsification studies~\cite{acrivos78,taylor64b, booty05,eggleton01,homsy04}. 
%The equation form is identical 
to the entrainment model for 
immiscible liquids in~\cite{wz04}, except that the surface 
tension contribution to $P_0$ is replaced here by the 
hydrostatic pressure difference. 

The natural boundary conditions are conditions on how the tendril 
shape $R(z)$ tapers to $0$ downstream as $z \rightarrow \infty$ 
and how $R(z)$ flares out into a flat interface upstream as 
$z \rightarrow 0$. Both conditions correspond to asymptotic 
balances of the governing equations. Far above the interface, 
the upwards plug flow $Ez$ dominates and 
\beq
R(z) \rightarrow R_\infty(z) = \sqrt{Q_0(c_0)/(\pi E z)} \quad {\rm as} \ z \rightarrow \infty, 
\label{infty}
\eeq
where we have written $Q_0(c_0)$ to emphasize that 
in our analysis, we use the scaling law (\ref{fluxlaw}) 
for $Q_0$, so that $c_0$ is the only undetermined parameter.
Near the tendril base, the effects of entrainment are 
negligible because the layers are strongly stratified.  The tendril 
base shape is therefore determined by a balance of the 
hydrostatic pressure and the exterior viscous stress, thus
\beq
R(z) \rightarrow R_s(z)= B \left(\frac{\ell_z}{z}\right)^{\alpha} e^{-{z}/{\ell_z}} \quad
{\rm as} \ z \rightarrow 0.
\label{upstream}
\eeq
Both the coefficient $B$ and the exponent $\alpha$ are
determined self-consistently with the full solution.  

\begin{figure}
\vspace{-10pt}
\psfrag{RRR}{\boldmath$R/\ell_{z}$}
\psfrag{zzz}{\boldmath$z/\ell_{z}$}
\includegraphics[width=0.3\textwidth, angle=270]{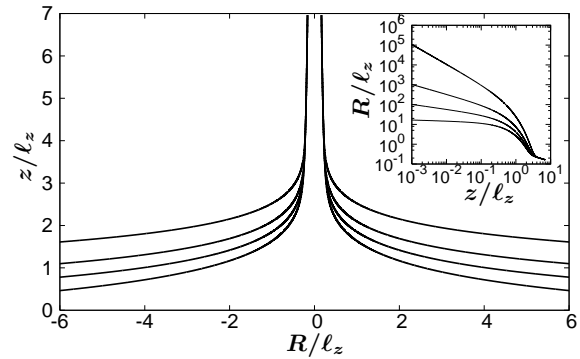}
\vspace{-3pt}
\caption{\label{family} 
Rescaled tendril solutions at 
$S/\ell_z=15$, $\mu_0/\mu=$~$0.1$, $c_0=3.1$ with exponents 
$\alpha=1.0,\, 0.5,\, 0.25,\, 0.052$ (top to bottom).
%$\alpha=-0.052, -0.25, -0.5$, and $-1.0$ (top to bottom). 
Inset shows tendril solution's power-law divergence.  }
\vspace{-15pt}
\end{figure}

Using boundary conditions (\ref{infty}), (\ref{upstream}), 
and appropriate downstream structural stability modes, 
we numerically integrated equations~(\ref{goveqn}) and 
(\ref{pressure}) for tendril solutions~\cite{schmidt}. 
We found that, for a fixed $Q_0$, a continuous family of 
tendril solutions exist. Each solution's upstream shape 
is characterized by a different set of $B, \alpha$ values. 
Fig.~\ref{family} gives a few examples of the different 
tendrils possible for the same volume flux.  The 
long-wavelength model is therefore clearly degenerate. 

Removing the degeneracy requires an additional condition 
on the upstream shape, one related to the large-scale 
topography. We next show this explicitly. In our toy model 
an axisymmetric withdrawal flow is generated in the upper 
layer by inserting a point force with strength $F$ at a 
height $S$ above the undisturbed interface. This is clearly 
not a realistic withdrawal near the point force but, if $S$ 
is large, it is a good approximation of realistic withdrawal 
flows near the interface. The velocity fields in both fluid layers
can be obtained by a variation of the method of 
images~\cite{lee79} and correspond to an axisymmetric straining 
flow with $E = F / (2 \pi \mu S^2)$ in the neighborhood of
the base of the tendril.  Since the effects of entrainment 
penetrate weakly into the lower layer, 
the large-scale interface shape should be 
well approximated by the deflection occurring when no 
liquid is entrained, 
\beq
{R_{\rm I}(z)} = S\ \sqrt{\left({3 \ell_z/2z}\right)^{2/5}-1} .
\label{outer}
\eeq 
The fact that $R_{\rm I}(z)$ is controlled by $S$ is the extra 
information needed to uniquely select a tendril solution.

\begin{figure}
\psfrag{lambda}[cc][cb]{\small{\boldmath$S/(\ell_{z}\sqrt{\mu_0/\mu})$}}
\psfrag{co}{\small{\boldmath$c_{0}$}}
\psfrag{R/Lz}{\large{\boldmath$R/\ell_{z}$}}
\psfrag{z/Lz}{\large{\boldmath$z/\ell_{z}$}}
\psfrag{key1}[rc][rc]{$R_{\rm I}$}
\psfrag{key2}[rc][rc]{$R(z,c_{0})$}
\includegraphics[width=0.25\textwidth, angle=270]{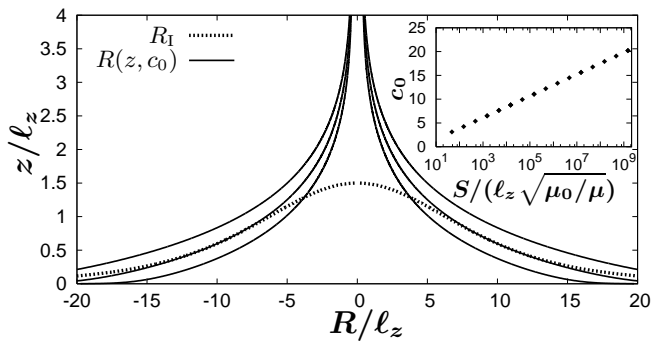}
\caption{\label{matchshape} 
Rescaled tendril solutions $R(z)$ for $c_0~=~3.4, 3.1$, and $2.9$
(solid lines, top to bottom), together with the large-scale deflection 
solution $R_{\rm I}(z)$ (dotted) for $S/\ell_z = 15$ and $\mu_0/\mu = 0.1$.
Only $c_0=3.1$ allows the two shapes to merge smoothly.
Inset shows the values of $c_0$ found from joining the numerical
solutions smoothly onto $R_{\rm I}(z)$ in this way. } 
\vspace{-15pt}
\end{figure}

To incorporate this extra information, we require that at an 
unknown location $z_s$, the tendril shape, as well as 
$\rd R/\rd z$ and $\rd^2 R/\rd z^2$ equal the corresponding 
quantities in $R_{\rm I}(z)$~\cite{fullstokes}.  We solve for 
$z_s$, $\alpha$ and $B$ analytically as follows. Since $z_s$ 
should be be near the base of the tendril, we approximate 
$R(z)$ by the upstream shape $R_s$~(\ref{upstream})
and find that $z_s \approx 0.609\ \ell_z$, $\alpha \approx 0.052$,
and $B \approx 1.18\ S$. 
We can solve for the appropriate tendril solution numerically 
by varying the dimensionless entrainment coefficient $c_0$ so 
that $R(z)$ and its derivative merge smoothly onto $R_{\rm I}(z)$.  
Fig.~\ref{matchshape} shows an example of how $c_0$ is found numerically. 

It is interesting to know how the dimensionless entrainment coefficient
$c_0$ depends on the global 
lengthscale, $S/\ell_z$, and on the viscosity contrast 
$\mu_0/\mu$.  We have varied $S/\ell_z$ from $10$ to $10^8$ 
and $\mu_0/\mu$ from $10^{-1}$ to $10^{-6}$ and plotted the 
results in the inset of Fig.~\ref{matchshape}. The variation 
of $c_0$ with these parameters is unexpectedly simple, well 
approximated by $c_0 = 2.3 \log \left( S/(\ell_{z}\sqrt{\mu_{0}/\mu}) \right) -0.74$.
The logarithmic dependencies of $c_0$ on $S/\ell_z$ and 
$\mu_0/\mu$ are weaker than the power-law dependencies of 
$Q_0$ on $\ell_z$ and $\mu_0/\mu$ in the scaling law 
(\ref{fluxlaw}). This shows that including the global 
dynamics only weakly modifies the simple entrainment law.

The logarithmic dependence ultimately results from the 
decaying exponential shape of the tendril at 
$z$~$\sim$~$\ell_z$~(\ref{upstream}).  How this emerges 
is complicated because $c_0$ is essentially
determined by demanding $R_\infty(z)$ merge 
onto $R_s(z)$ through a region where all the terms in 
the governing equation~(\ref{goveqn}) are equally important. Our rough 
analysis shows that two requirements appear necessary 
for the observed $c_0$ relation. First, the upstream 
tendril shape $R_s(z)$ should assume a size consistent with 
$R_\infty(z)$, the steady-state downstream shape.  
Second, the tapering of the tendril must be accomplished 
by $z \approx \ell_z$, with the sharp exponential decay 
in $ \rd R_s(z)/\rd z$ appropriately switching over to a 
gentler square-root decay in $\rd R_\infty(z)/\rd z$~\cite{schmidt}.  

In sum, we have now an improved scaling law for $Q_0$, one 
which explicitly accounts for the dependencies on the 
global flow parameters
\beq
Q_0 = \frac{16 \pi \mu^2 \mu_0 E^4}{(\Delta \rho g)^3}\left(\gamma_1 \log \left( \frac{S}{\ell_{z}\sqrt{\mu_{0}/\mu}} \right) +\gamma_2 \right),  
\label{sol}
\eeq
where $\gamma_1=2.3$ and $\gamma_2=-0.74$. 
We have also analyzed tendril solutions associated with 
the withdrawal flows generated by a ring vortex and by 
a point sink and found the same qualitative outcome. 
The differences in geometry between these withdrawal 
flows result in slightly different tendril shapes and 
values of $\gamma_1$ and $\gamma_2$.  Using (\ref{sol}) 
we have estimated how quickly two miscible layers will 
mix in thermal convection and found results consistent 
with observed values. However, the strong dependence on 
$E$ of (\ref{sol}) makes a precise comparison difficult.  
To fully establish how the tendril persists over time, 
a complete numerical study as well as an experiment 
where the withdrawal is generated directly, e.g. by 
withdrawing liquid from a tube inserted into the upper 
layer~\cite{cohen02,courrech06,case07}, instead of 
indirectly via thermal convection, are needed.  

Aside from providing an estimate for $Q_0$ that takes 
into account the interior stagnation point, our analysis 
also reveals that, although some information about the 
large-scale flow is necessary to specify a tendril 
shape, the dependence on the large-scale flow geometry 
is weak.  This may be why tendrils in thermal convection 
experiments can remain stable over long periods of time 
despite fluctuations in the global convection which can 
alter the large-scale topography.  

\begin{acknowledgments}
The authors thank Anne Davaille, Sarah C. Case, Leo P. Kadanoff and 
Sidney R. Nagel for encouragement and helpful discussions. 
This work was supported by a GAANN Fellowship (L.E.S.) and 
NSF MRSEC DMR-0213745.
\end{acknowledgments}
\vspace{-10pt}
\bibliography{tendrilcite}

\begin{thebibliography}{31}
\bibitem[{ful()}]{fullstokes}
\bibinfo{note}{Here we have chosen the simpler option of matching $R_{\rm
  I}(z)$ and $R(z)$ at a point instead of systematically matching them. Our
  numerics show the tendril merges smoothly onto $R_{\rm I}(z)$, so the errors
  appear small. The simpler procedure also has the advantage of revealing
  parameter dependencies in an explicit way.}
\end{thebibliography}
\end{document}